# Optical bistability of continuous-wave and multi-pulse phase transition within EDFL via low threshold saturable absorber

Hsuan-Sen Wang,[1] Wen-Hsuan Kuan,[4, 5] Ahmed F. M. El-Mahdy,[2] Gong-Ru Lin,[3] Kuei-Huei Lin,[4] Shiao-Wei Kuo,[2, 6] and Chao-Kuei Lee,[*, 1, 5]

[1] *Department of Photonics, National Sun Yat-Sen University, Kaohsiung 80424, Taiwan*
[2] *Department of Materials and Optoelectronic Science, National Sun Yat-Sen University, Kaohsiung 80424, Taiwan*
[3] *Graduate Institute of Photonics and Optoelectronics and Department of Electrical Engineering, National Taiwan University, Taipei 10617, Taiwan*
[4] *Department of Applied Physics and Chemistry, University of Taipei, 1, Ai-Guo West Road, Taipei 10048, Taiwan*
[5] *These authors contributed equally to this work.*
[6] *kuosw@faculty.nsysu.edu.tw*
*\*chuckcklee@yahoo.com*

**We demonstrate optical bistability in an erbium-doped fiber laser (EDFL) using a low saturation intensity covalent organic framework (COF) saturable absorber (SA). The COF-SA satisfies the free-energy criterion for bistability, enabling switching among non-lasing, continuous-wave, and mode-locking states. Two optical bistability regions are observed, including a distinctive direct mode-locking to non-lasing transition during pump down-sweep, absent in previous EDFL studies. Stable soliton-like pulses and stepwise pulse-number hysteresis further indicate first-order phase-transition-like dynamics. These results show that low-saturation intensity SAs offer a compact route to bistability-assisted pulse formation in fiber lasers.**

Optical bistability (OB) is a phase transition between two distinct optical responses that occur under a single-input condition, such as two levels of reflectivity, transmission, or refractive index at the same power and wavelength. This dual-state behavior is characterized by a hysteresis loop in the input–output relation, where the output depends on the excitation history [1,2]. The ability to sustain two stable outputs under identical conditions makes OB attractive for applications in all-optical switching and optical memory, driving extensive interest [3]. Beyond these applications, observing OB provides a sensitive probe of nonlinear light–matter interactions and the boundaries between different dynamical regimes, while its well-defined thresholds offer stringent benchmarks for validating nonlinear photonic models, in particular within compact fiber-laser platforms.

According to earlier predictions [4,5], a laser incorporating a saturable absorber (SA) can exhibit OB, and this behavior has been experimentally confirmed in several erbium-doped fiber laser (EDFL) configurations [6–10]. OB has been demonstrated using an unpumped segment of erbium-doped fiber (EDF) acting as an effective SA or by employing a sufficiently long gain fiber in which pump depletion produces a partially unpumped fiber tail. The intracavity signal can bleach this tail, reducing loss and providing the intensity-dependent transmission required for bistability [7]. In these studies, stable continuous-wave (CW) lasing was obtained above pump threshold, and more recently, self-pulsing behavior has also been observed in ring-cavity EDFL [11]. This phenomenon was attributed to the combined effects of ion-pair interactions and reabsorption, with experimental observations supported by coupled-rate-equation simulations. These developments highlight the rich nonlinear dynamics accessible in such systems and motivate further exploration of additional types of OB in lasers with saturable absorption. In most of laser systems, either off–to–CW bistability (spontaneous emission ↔ CW) or CW–to–pulsation bistability (CW ↔ pulsed) was reported. However, the coexistence of both types of OB in one laser configuration has not been observed.

Therefore, we propose replacing the unpumped lengthy EDF segment with a compact SA, which is conventionally used for generating optical pulses in the laser cavity, to provide the nonlinearity required for emerging new types of OB. The foundation for this approach lies in Scott and Scully's theories, which drew a parallel between Landau phase transitions and laser dynamics near the threshold, and linked OB with low-threshold SA to thermodynamic phase transitions [5,12].

When an SA is introduced into the laser system, the loss expression incorporating this effect can be described as $\sigma_t(E^2) = \sigma_{t0} + \sigma_s/(1 + E^2/I_s)$, where $E$ is the electric field, $I_s$ and $\sigma_s$ are saturation intensity and linear absorbance of the saturable absorber, and $\sigma_{t0}$ is the cavity loss of the laser cavity. As a result, the Gibbs free energy has to be revised as

$$G(E) = -\tfrac{1}{2}A[(\sigma - \sigma_{t0})E^2 - \sigma_s I_s \log(1 + E^2/I_s)] + \tfrac{1}{4}B\sigma E^4 \quad (1)$$

This modified free-energy expression provides a direct way to analyze how the SA reshapes the intracavity energy landscape, since the nonlinear loss term introduced by the SA alters the curvature of G(E) and determines the number and stability of its minima. In the low-field limit, the Taylor expansion of Eq. (1) indicates that while $I_s \leq A\sigma_s/(B\sigma)$, the free energy profile changes from Fig. 1(a) to Fig. 1(b). For the latter, the positive electric field signatures the order parameter in the first-order phase transition [5]. In this regime, the free-energy landscape develops two competing minima, corresponding to a low-intensity (non-lasing) state and a high-intensity (lasing) state, which coexist over a finite pump range. Namely, the laser system exhibiting bistable operation can be anticipated when the laser system is modulated a SA with low saturation intensity, which may provide the theoretical guideline for generating OB in a laser system.

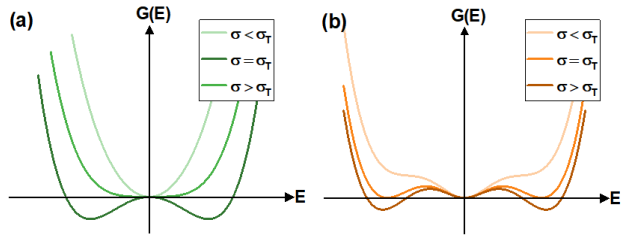

Fig. 1. Free-energy G(E) profile from Eq. (1): (a) $I_s > A\sigma_s/B\sigma$, showing a single-well (second-order) transition; (b) $I_s < A\sigma_s/B\sigma$, showing a double-well (first-order) transition.

In this study, we explore OB in an EDFL by incorporating a porphyrin/pyrene-linked COF as the saturable absorber [13]. The COF-SA provides a low saturation threshold together with robust thermal and chemical stability, making it a promising candidate for achieving the low–$I_s$ condition identified in the free-energy analysis. By integrating this SA into a short EDF cavity, we examine whether the resulting nonlinear loss can generate the predicted double-well landscape and support multiple coexisting operating states within a single configuration. We also aim to determine whether bistable switching may occur beyond the conventional CW–pulsed scenario and whether the COF-SA influences multipulse dynamics in a manner consistent with phase-transition-like behavior. Through this approach, we seek to clarify how a low-saturation-intensity SA can guide OB and potentially enable new pulse-formation pathways in fiber-laser systems.

The experimental setup is schematically depicted in Fig. 2. The laser is configured as a unidirectional ring cavity. The gain medium is a 2 m-long EDF with an absorption coefficient of 80 dB/m at 1531 nm and a dispersion parameter of 24.6 ps$^2$/km. The cavity also contains 106 m of single-mode fiber with a dispersion parameter of 23.6 ps$^2$/km. The total cavity length is approximately 108 m, giving a fundamental repetition rate of 1.85 MHz. A 976 nm laser diode serves as the pump source. The cavity incorporates a 980/1550 nm wavelength-division multiplexer (WDM), a 95:5 output coupler (OC), a polarization-independent optical isolator (PI-ISO), and a COF-PVA SA positioned between two FC/PC fiber connectors. As reported in our previous work [13], the COF-PVA SA exhibits a modulation depth of approximately 0.8%, a saturation intensity of 0.5 kW/cm$^2$, and a nonsaturable loss of 84.2%, the latter mainly arising from coupling loss due to the thickness of the PVA film.

To investigate the laser output characteristics, the pump current was adjusted in 10 mA steps between 300 and 400 mA. After each adjustment, the output was monitored for approximately one minute until the laser state remained unchanged to ensure steady-state operation before measurement. The laser output was extracted from the 5% port of the OC during both increasing and decreasing pump scans. The average power, optical spectrum, and temporal waveforms were simultaneously monitored using a power meter (Newport, Model 1918-R), an optical spectrum analyzer (Anritsu, MS9740B), and a real-time oscilloscope (Tektronix, MDO4104B-6), respectively.

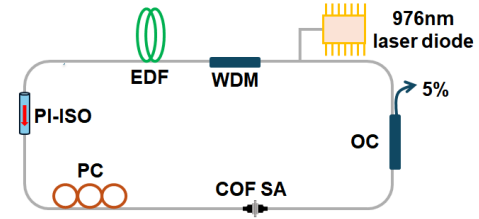

Fig. 2. Experimental setup of the COF-SA-based erbium-doped fiber laser.

The power bistability of the laser is evident from the hysteresis loop in the output-power–versus–pump-current curve spanning 310–380 mA, as shown in Fig. 3. When the pump current is increased (blue markers), the laser reaches threshold and begins CW operation at 350 mA. CW lasing persists until the pump is raised to 380 mA, at which point the laser switches to a stable mode-locked state. In contrast, when the pump current is decreased (green markers), the mode-locked state remains stable down to 310 mA, below which the laser abruptly returns to the non-lasing state.

Two distinct types of OB appear in the full hysteresis curve. The first occurs between 310 and 350 mA, where non-lasing and mode-locked states coexist. The second appears between 350 and 380 mA, where CW and mode-locked states coexist. As a result, within the broad pump interval of 310–380 mA, the laser may operate in non-lasing, CW, or mode-locked states, with the accessible state determined entirely by the history of the pump sweep.

In typical fiber lasers, the pump-up trajectory follows the sequence non-lasing → CW → mode locking, with the reverse sequence observed during pump-down. In contrast, our system exhibits a direct mode-locking → non-lasing transition during the pump reduction, bypassing the intermediate CW state entirely. This form of bistability provides a practical advantage for generating high-contrast optical pulses, as the pulsed state remains robust over a wider pump range before collapsing directly into the off state.

Figures 4 and 5 present the optical spectra and temporal traces of the EDFL output, respectively. The analysis begins with the increasing pump scan, corresponding to points (1) and (2) in Fig. 3. At 370 mA (point 1), the laser operates in the CW regime, producing a sharp spectrum centered at 1570 nm (curve 1 in Fig. 4). When the pump is increased to 380 mA (point 2), a dramatic spectral transformation occurs: the spectrum broadens to a 3-dB bandwidth of 2.6 nm, the central wavelength redshifts to 1573.6 nm, and distinct symmetric Kelly sidebands appear, confirming the onset of stable mode locking.

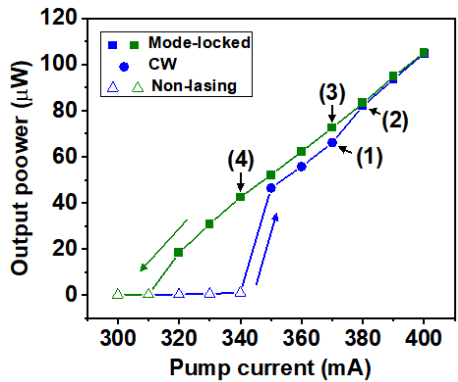

Fig. 3. Output power versus pump current for increasing and decreasing pump sweeps in the COF-SA-based EDFL.

The bistable behavior becomes evident in the decreasing pump scan (points 3 and 4). At 370 mA (point 3), despite having the same pump current as point (1), the laser does not revert to the CW state. Instead, its spectrum remains broad and nearly identical to the mode-locked spectrum at point (2). This mode-locked signature persists even as the pump is reduced to 340 mA (point 4). Thus, OB is clearly demonstrated in the spectral domain: the same pump current can yield two distinct spectral states, determined entirely by the pumping history.

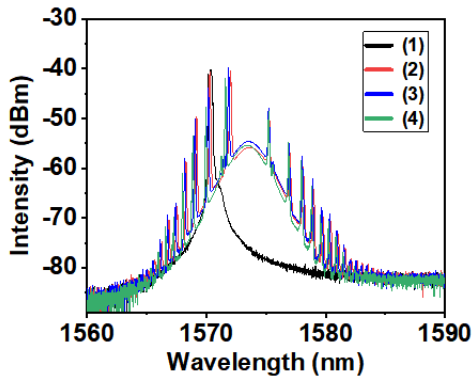

Fig.4. Optical spectra for states (1)–(4) identified in Fig. 3 under the corresponding pump-scan directions.

Figure 5 shows the time-domain waveforms of the laser output, providing direct evidence of the operational states corresponding to the points labeled in Fig. 3. During the increasing pump scan, the black trace (1) recorded at 370 mA exhibits a constant intensity level with no pulsation, characteristic of CW operation. In contrast, the red trace (2), obtained at 380 mA, displays a stable train of well-defined periodic pulses, confirming the transition to mode locking. The decreasing pump scan further supports this bistability: at the same pump current of 370 mA, the blue trace (3) remains in the mode-locked state rather than returning to the CW state observed in trace (1). As the pump is reduced to 340 mA, the green trace (4) shows that pulsed operation persists. Together, these oscilloscope traces clearly demonstrate OB between CW and mode-locked states. Moreover, the differences in pulse spacing and pulse number among traces (2)–(4) indicate phase-transition-like behavior in the evolution of the mode-locked states under varying pump conditions.

The state-transition diagram in Fig. 6 effectively illustrates the laser's operational paths under increasing and decreasing pump currents. It highlights how our system follows different trajectories depending on its excitation history, revealing the presence of two distinct bistable regions associated with the non-lasing/CW and CW/mode-locking transitions. This switching pathway reflects the underlying first-order, double-well free-energy landscape discussed earlier, in which the intracavity field jumps between coexisting stable states once the pump crosses the corresponding energy barrier. The diagram therefore provides a concise representation of the bistable dynamics in our laser cavity and helps clarify how the COF-SA influences the transitions between the various steady states.

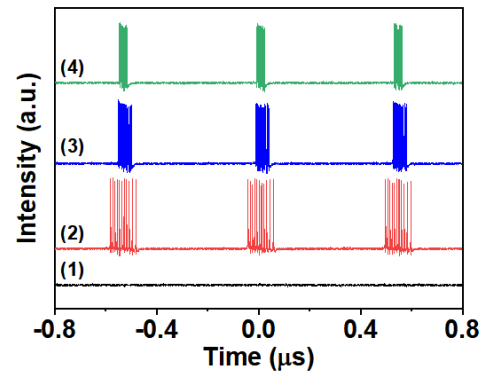

Fig. 5. Temporal traces for states (1)–(4) indicated in Fig. 3 under the corresponding pump-scan directions.

Although previous studies on OB in fiber lasers did not explicitly insert a separate SA into the cavity, it was recognized that a sufficiently long EDF can itself give rise to bistability [5,6]. A long EDF effectively behaves as two segments: the pumped section provides gain by absorbing pump photons, while the unpumped tail acts as a saturable absorber. In contrast, by combining a short EDF with a COF-SA, we observe a clear hysteresis loop in the output–pump diagram and confirm the appearance of OB in the low-power region. These observations are consistent with the laser–phase-transition analogy proposed by Scott et al., which predicts that bistability arises when the cavity incorporates an SA with sufficiently low saturation intensity.

However, pulse formation was not reported in Refs. [5–6] because the effective SA provided by the unpumped EDF tail possesses a relatively long relaxation time. In addition, the long EDF introduces net normal dispersion in the cavity, which suppresses the formation of optical solitons or soliton bunches. These studies therefore provide useful contrast: in our cavity, the much shorter SA relaxation time together with net anomalous dispersion—arising from the COF-SA and short EDF—creates conditions favorable for soliton-like pulse formation and dynamic pulse bunching.

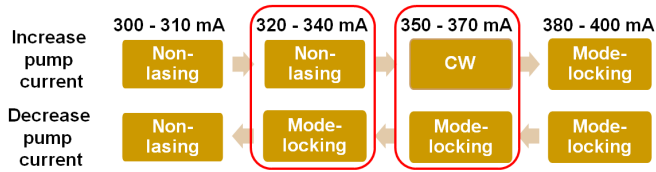

Fig. 6. Path-dependent laser-state transitions showing bistability between non-lasing and mode-locked states (320–340 mA) and between CW and mode-locked states (350–370 mA). Arrows indicate dependence on pump-sweep direction.

In addition to the pump-power hysteresis, a pronounced pulse-number hysteresis is also observed in our system. Unlike the behavior reported in [14], where multiple pulses emerge or vanish strictly one at a time as the pump strength is varied, our cavity exhibits discrete, step-like changes in pulse number even under fine pump-current adjustments. This discontinuous evolution arises from the free-energy landscape imposed by the low-saturation-intensity COF-SA: the strong nonlinear loss reshapes the intracavity energy profile into multiple metastable minima. When the pump current crosses specific thresholds, the pulse ensemble transitions collectively between these free-energy minima, causing groups of pulses to appear or disappear simultaneously rather than following a smooth, pulse-by-pulse sequence. As a result, the system exhibits a first-order, phase-transition-like reorganization of intracavity states, manifested experimentally as the observed stepwise pulse-number hysteresis.

Beyond demonstrating COF-SA–induced bistability, our results highlight a broader design principle for achieving stable mode locking in EDFLs. Recent theoretical studies have shown that dissipative soliton formation in nonlinear resonators is strongly favored when the system possesses an intrinsic bistable response [15]. In our work, the low-saturation-intensity COF-SA reshapes the intracavity dynamics into a coexistence region where localized, soliton-like pulses can condense and remain robust. Thus, the stable mode-locked states observed in our system arise not only from anomalous dispersion but from a bistable energy landscape that supports pulse localization, offering a practical route toward compact, highly stable mode-locked fiber lasers.

From the perspective of the free-energy landscape and first-order phase-transition theory, we have shown that incorporating a low-saturation-intensity COF-based SA reshapes the intracavity dynamics of an EDFL and enables OB. Experimentally, we observe bistable transitions not only between CW and mode-locking states but also between non-lasing and mode-locking states, allowing a direct transition from pulsed operation to the off state during the pump down-sweep. Such a transition has not been reported in previous EDFL studies, which consistently exhibited only the sequential non-lasing → CW → mode-locking path and its reverse.

This direct collapse of the pulsed state can be understood as a first-order, discontinuous jump between free-energy minima: as the pump decreases, the effective gain–loss balance can no longer sustain the high-intensity minimum, and the COF-SA rapidly unbleaches, sharply increasing the nonlinear loss. The system therefore abandons the pulsed minimum and falls directly into the zero-intensity minimum without passing through the CW state.

Beyond explaining this unconventional state transition, the bistable landscape created by the COF-SA establishes a coexistence region that strongly promotes the formation of stable, soliton-like pulses—consistent with recent theoretical insights linking dissipative soliton formation to bistability in nonlinear optical resonators. Although the COF-SA parameters were not fully optimized, our results demonstrate that low-saturation-intensity SAs offer a powerful and generalizable route to bistability-assisted pulse generation, providing practical guidelines for compact EDFLs capable of high-stability mode locking and new forms of state switching not accessible in traditional fiber-laser architectures.

**Funding.** National Science and Tecnology Council (113-2221-E-845-003,114-2221-E-845-002)

**Disclosures.** The authors declare no conflict of interest.

**Data availability.** Data underlying the results presented in this paper are not publicly available at this time but may be obtained from the authors upon reasonable request.

**Reference**
1. H. M. Gibbs, (Academic Press, 1985), pp. 19–92.
2. J. M. H. Barakat, A. S. Karar, R. Ghandour, and Z. N. Gürkan, Result Eng. **26**, 105540 (2025).
3. J. Chen and X. Liu, Nat. Photonics **19**, 122 (2025).
4. P. Casagrande and L. A. Lugiato, Nuov Cim B **48**, 287 (1978).
5. J. F. Scott, M. Sargent, and C. D. Cantrell, Opt. Commun. **15**, 13 (1975).
6. J. C. Martin, IEEE Photon. Technol. Lett. **28**, 292 (2015).
7. J. Shao, S. Li, Q. Shen, Z. Wu, Z. Cao, and J. Gu, Opt. Express **15**, 3673 (2007).
8. J. M. Oh and D. Lee, IEEE J. Quantum Electron. **40**, 374 (2004).
9. S. Li, Q. Ge, Z. Wang, J. C. Martín, and B. Yu, Sci Rep **7**, 8992 (2017).
10. Q. Ge, S. Li, Z. Wang, S. Zhen, J. C. Martín, and B. Yu, Opt. Laser Technol. **98**, 79 (2017).
11. J. Shang, T. Feng, S. Zhao, J. Zhao, Y. Zhao, Y. Song, and T. Li, Appl. Phys. Express **13**, 112006 (2020).
12. V. DeGiorgio and M. O. Scully, Phys. Rev. A **2**, 1170 (1970).
13. H.-S. Wang, A. F. M. EL-Mahdy, S.-W. Kuo, G.-R. Lin, and C.-K. Lee, Chin. J. Phys. **89**, 964 (2024).
14. X. Liu, Phys. Rev. A **81**, 023811 (2010).
15. L. Lugiato, F. Prati, M. Brambilla, and L. L. Columbo, Nanophotonics **14**, 3459 (2025).